\documentclass[12pt]{article}
\begin{document}
\title{Lie Point Symmetries for Reduced Ermakov Systems}
\author{
F.~Haas\footnote{ferhaas@exatas.unisinos.br} \,\,and J.~Goedert\\
Centro de Ci\^encias Exatas e Tecnol\'ogicas -- UNISINOS \\
Av. Unisinos, 950\\
93022-000 S\~ao Leopoldo, RS, Brazil
}
\maketitle
\begin{abstract}
\noindent Reduced Ermakov systems are defined as Ermakov systems
restricted to the level  surfaces of the Ermakov invariant. The
condition for Lie point symmetries for reduced Ermakov systems is
solved yielding four infinite families of systems. It is shown
that $SL(2,R)$ always is a group of point symmetries for the
reduced Ermakov systems. The theory is applied to a model example
and to the equations of motion of an ion under a generalized Paul
trap.
\end{abstract}

\noindent
\hspace{.8cm} {\it PACS numbers:} 02.30.Hg; 02.90.+p; 03.20.+i\\
\vspace{-.5cm}

\hspace{.2cm} {\it Keywords:} Ermakov system; Lie symmetry; Ermakov invariant.

\section{Introduction}
Ermakov systems \cite{ermakov}-\cite{lewis} play an important role
in a variety of physical and mathematical situations. The most
recent analyses involving Ermakov systems deal with Bose-Einstein
condensates and cosmological models \cite{haasPRA}-\cite{hawkins},
nonlinear supersymmetric Darboux transformations \cite{ioffe}, the
free fall of a quantum bouncing ball \cite{rosu}, conformal
quantum mechanics \cite{akulov} and generalized Hamiltonian
structures \cite{haasJPA}. From the theoretical viewpoint, Ermakov
systems always admit a constant of motion, the Ermakov invariant,
and are amenable to a nonlinear superposition law \cite{reid}. In
addition, Ermakov systems are linearizable under broad
circumstances \cite{athornelinear, haasg}.

As is well known, the group theoretic analysis of a dynamical
system is a subject of great relevance \cite{olver} not only for
the reduction of order and the search for invariants for the
system but also for a better understanding of its structural
properties. The point symmetry group of Ermakov systems has been
identified as the $SL(2,R)$ group \cite{leach}-\cite{goedert}.
More recently, using the converse to Noether's theorem, it has
been shown that the Ermakov invariant can be associated to a
dynamical symmetry, in the cases where the Ermakov system admits a
variational formulation \cite{haasPLA}. In addition, $SL(2,R)$ has
also been found \cite{karasu} as the symmetry group of
Kepler-Ermakov systems \cite{athornel}, which can be viewed either
as perturbations of the planar Kepler problem or of the classical
Ermakov system. The purpose of this paper is to follow this trend
from a different perspective and study the Lie point symmetries of
Ermakov systems restricted to manifolds where the Ermakov
invariant has a fixed constant value. The importance of this study
may not be underestimated since the existence of the Ermakov
invariant is automatic.

In polar coordinates, the Ermakov system reads
\begin{eqnarray}
\label{e1}
\ddot{r} - r{\dot\theta}^2 &+& \omega^{2}r = \frac{F(\theta)}{r^3} \,,\\
\label{e2}
r\ddot\theta + 2\dot{r}\dot\theta &=& \frac{G(\theta)}{r^3} \,,
\end{eqnarray}
where $F$ and $G$ are arbitrary functions of the angle and
$\omega$, in principle, can depend arbitrarily on the dynamical variables.
More often, $\omega$ is a function of time only, in which case it
has the interpretation of a time-dependent frequency. For
simplicity, we consider the case $\partial\omega/\partial\dot{r} =
0$. No mater the special form of $\omega$, the Ermakov systems
always possess the constant of motion
\begin{equation}
\label{e3}
I = \frac{1}{2}(r^{2}\dot\theta)^2 - \int^{\theta}G(\phi)\,d\phi \,,
\end{equation}
the so-called Ermakov invariant \cite{ermakov}-\cite{lewis}.

The existence of a constant of motion allows the reduction of the
order of the system. More exactly, the fourth-order system
(\ref{e1}-\ref{e2}) can be rewritten as
\begin{eqnarray}
\label{red1}
\ddot r &=& A(r,\theta,t) \,,\\
\label{red2}
r^{2}\dot\theta &=& B(\theta) \,,
\end{eqnarray}
where the functions $A$ and $B$ are defined according to
\begin{eqnarray}
\label{e4}
A(r,\theta,t) &=& - \omega^{2}r + \frac{1}{r^3}(F(\theta) + 2I + 2\int^{\theta}G(\phi)\,d\phi) \,,\\
\label{e5}
B(\theta) &=& \sqrt{2}(I + \int^{\theta}G(\phi)\,d\phi)^{1/2} \,.
\end{eqnarray}
The fact that $\partial\omega/\partial\dot{r} = 0$ ensures the
indicated functional dependence of $A$ in (\ref{e4}). Notice,
however, that $\omega$ can freely depend, for instance, on
$\dot\theta$, since this dependence can be eliminated through
(\ref{red2}). The choice $\partial\omega/\partial\dot{r} = 0$ has
a decisive influence on the simplification of the symmetry
analysis so that we do not claim our results are the most general.

Equations (\ref{red1}-\ref{red2}) are a third-order dynamical
system which we call the {\it reduced Ermakov system}. Even if we
do not show explicitly, notice that the reduced equations do
depend parametrically on $I$. Also, the function $B$ is not
identically zero except in the trivial case $I = G(\theta) = 0$,
which we do not consider here.

The purpose of this work is to perform the Lie point symmetry
analysis of reduced Ermakov systems. Since for Lagrangian Ermakov
systems the Ermakov invariant is directly related to a dynamical
Noether symmetry \cite{haasPLA}, it is to be expected that the
algebra $sl(2,R)$ do play a fundamental role on the reduced
Ermakov system.

The organization of the paper is as follows. In Section 2, the
general symmetry conditions to be satisfied by reduced Ermakov
systems and his generator of symmetries are determined. In Section
3, we solve the symmetry conditions in the case of transformations
of the time parameter not involving the dynamical coordinates.
This yields four classes of reduced Ermakov systems admitting Lie
point symmetries. In Section 4, we show that, irrespective of its
specific form, the reduced Ermakov systems always do admit
$SL(2,R)$ as a symmetry group. This shows that, more properly,
$SL(2,R)$ is the group of symmetries for {\it reduced} Ermakov
systems, the reduction being possible thanks to a dynamical
Noether symmetry in the case of Lagrangian Ermakov systems. In
Section 5, we apply the whole formalism to a particular example,
looking for reduced Ermakov systems admitting a chosen Lie
symmetry. In Section 6, we start from the equations for an ion
under a generalized Paul trap and find the circumstances under
which these equations, written as a reduced Ermakov system, do
possess Lie point symmetries. Section 7 is devoted to the
conclusions.

\section{Lie Symmetries}
Consider infinitesimal point transformations of the form
\begin{eqnarray}
\label{e6}
\bar{r} &=& r + \varepsilon R(r,\theta,t) \,,\\
\label{e7}
\bar{\theta} &=& \theta + \varepsilon S(r,\theta,t) \,, \\
\label{e8}
\bar{t} &=& t + \varepsilon T(r,\theta,t) \,,
\end{eqnarray}
for functions $R$, $S$ and $T$ to be determined and infinitesimal
parameter $\varepsilon$.  The procedure for computing Lie
symmetries is well known \cite{olver} and we limit ourselves to
sketch the critical steps in our case. The above infinitesimal
transformation will be a Lie point symmetry of the reduced Ermakov
system if and only if (\ref{red1}-\ref{red2}) remains formally
invariant under (\ref{e6}-\ref{e8}) up to first order in
$\varepsilon$, in the solution set of the reduced Ermakov system.
This symmetry condition will imply the vanishing of two separate
polynomials in $\dot{r}$, one associated with the radial equation
(\ref{red1}), the other associated with the angular equation
(\ref{red2}). Assuming the vanishing of the coefficient of
different powers of $\dot{r}$ we get the following set of linear,
coupled partial differential equations,
\begin{eqnarray}
\label{sc1}
\frac{\partial^{2}T}{\partial r^2} &=& 0 \,, \\
\label{sc2}
\frac{\partial^{2}R}{\partial r^2} &-& \frac{2B}{r^2}\frac{\partial^{2}T}{\partial r\partial\theta} - 2\frac{\partial^{2}T}{\partial r\partial t} + \frac{2B}{r^3}\frac{\partial T}{\partial\theta} = 0 \,,\\
\label{sc3}
\frac{2B}{r^2}\frac{\partial^{2}R}{\partial r\partial\theta} &+& 2\frac{\partial^{2}R}{\partial r\partial t} - \frac{2B}{r^3}\frac{\partial R}{\partial\theta} - 3A\frac{\partial T}{\partial r} - \frac{BB'}{r^4}\frac{\partial T}{\partial\theta} \\ &-& \frac{B^2}{r^4}\frac{\partial^{2}T}{\partial\theta^2} - \frac{2B}{r^2}\frac{\partial^{2}T}{\partial\theta\partial t} - \frac{\partial^{2}T}{\partial t^2} = 0 \,, \nonumber \\
\label{sc4}
r^{2}\frac{\partial S}{\partial r} &-& B\frac{\partial T}{\partial r} = 0 \,,\\
\label{sc5}
B\frac{\partial S}{\partial\theta} &-& S\,B' + r^{2}\frac{\partial S}{\partial t} - \frac{B^2}{r^2}\frac{\partial T}{\partial\theta} - B\frac{\partial T}{\partial t} + \frac{2RB}{r} = 0 \,,\\
\label{sc6}
R\frac{\partial A}{\partial r} &+& S\frac{\partial A}{\partial\theta} + T\frac{\partial A}{\partial t} = \left(\frac{\partial R}{\partial r} - \frac{2B}{r^2}\frac{\partial T}{\partial\theta} - 2\frac{\partial T}{\partial t}\right)\,A  \\
&+& \frac{B^2}{r^4}\frac{\partial^{2}R}{\partial\theta^2} + \frac{BB'}{r^4}\frac{\partial R}{\partial\theta} + \frac{2B}{r^2}\frac{\partial^{2}R}{\partial\theta\partial t} + \frac{\partial^{2}R}{\partial t^2} \,. \nonumber
\end{eqnarray}
In equations (\ref{sc3}) and (\ref{sc6}) and in the sequel, a
prime denotes total derivative  with respect to $\theta$, so that
$B' = dB/d\theta$. The solutions of the equations
(\ref{sc1}-\ref{sc6}) determine the Lie point symmetries of the
reduced Ermakov system as well as the classes of admissible
functions $A$ and $B$. In the following Section we show four
categories of solutions for the determining equations.

\section{Exact solutions}
A closer examination of (\ref{sc3}) shows that the function $A$
which specifies the dynamics of the radial variable in the reduced
Ermakov system, will soon become, to some extent, determined if
$\partial T/\partial r \neq 0$. Indeed, (\ref{sc1}) and
(\ref{sc2}) immediately give the $r$ dependence of $R$ and $T$,
which, in turn, will determine the $r$ dependence of $A$ through
(\ref{sc3}). Furthermore, using (\ref{sc4}-\ref{sc5}) shows that,
for $\partial T/\partial r \neq 0$, $A$ contains only two terms,
one proportional to $r$, the other to $r^{-3}$. To avoid this
excessively constrained situation we must have
\begin{equation}
\label{e9}
\frac{\partial T}{\partial r} = 0 \,,
\end{equation}
a condition assumed throughout this paper.

If (\ref{e9}) is valid, the solution for (\ref{sc1}-\ref{sc5}) is
\begin{eqnarray}
\label{e10}
R &=& (\rho(t)\dot\rho(t) +  \Gamma(\theta))\,r \,,\\
\label{e11}
S &=& S(\theta) = \left(\kappa - 2\int^{\theta}d\phi\,\frac{\Gamma(\phi)}{B(\phi)}\right)\,B(\phi)     \,,\\
\label{e12}
T &=& \rho^{2}(t) \,,
\end{eqnarray}
where $\rho$ is an arbitrary function of time, $\Gamma$ is an
arbitrary function of $\theta$ and $\kappa$ a numerical constant.

Until now, no constraint was imposed on the functions $A$ or $B$
of the reduced Ermakov system, but there still remains the
symmetry condition (\ref{sc6}). Inserting (\ref{e10}-\ref{e12})
into (\ref{sc6}), we get the following determining equation,
\begin{equation}
\label{d} U\,A = (-3\rho\dot\rho + \Gamma)\,A +
(\rho{\buildrel\cdots\over\rho} + 3\dot\rho\ddot\rho)\,r +
\frac{B}{r^3}(B\Gamma'' + B'\Gamma') \,,
\end{equation}
where $U$ is the generator of Lie point symmetries for reduced Ermakov systems,
\begin{equation}
\label{e13}
U = (\rho\dot\rho + \Gamma)\,r\frac{\partial}{\partial r} + S\frac{\partial}{\partial\theta} + \rho^{2}\frac{\partial}{\partial t} \,.
\end{equation}
The generator $U$, contains two arbitrary functions, $\rho(t)$ and
$\Gamma(\theta)$, and one arbitrary numerical constant, $\kappa$,
implicit in the definition (\ref{e11}) of $S(\theta)$.

The $sl(2,R)$ algebra is obtained from (\ref{e13}) in the particular case
\begin{equation}
\label{e14}
\rho^2 = c_0 + c_{1}t + c_{2}t^2 \,, \quad \Gamma = \kappa = 0 \,,
\end{equation}
where $c_0$, $c_1$ and $c_2$ are arbitrary numerical constants.
The three generators of the $sl(2,R)$ algebra is obtained by
taking separately each of these constants non zero. In comparison
with the generator of point symmetries of the non reduced Ermakov
system \cite{leach}, the new ingredients of $U$ are
$\Gamma(\theta)$ and $\kappa$. Also, $\rho^{2}(t)$ is not
necessarily a second degree polynomial in $t$. In Section 4, we
present a more detailed account of the relation between the point
symmetries of the reduced Ermakov systems and the $sl(2,R)$
algebra.

Equation (\ref{d}) can be viewed either as an equation for $A$ or
for $B$. We feel it is more productive to think of it as a
determining equation for $A$, since $B$ participates in the
generator through the definition (\ref{e11}). Following this
choice, we find four classes of solutions for $A$ satisfying
(\ref{d}), listed bellow. All these solutions are build using the
differential invariants of the operator $U$, that is, the
independent functions $I_1$ and $I_2$ for which $U(I_{1}) =
U(I_{2}) = 0$.

\subsection{The $\rho \neq 0$, $S \neq 0$ Case}
In this situation, the method of characteristics yields the
following differential invariants for the generator $U$,
\begin{eqnarray}
\label{e16}
I_1 &=& \int^{\theta}\frac{d\phi}{S(\phi)} - \int^{t}\frac{d\tau}{\rho^{2}(\tau)} \,,\\
\label{e17}
I_2 &=& \frac{r}{\rho}\exp\left(-\int^{t}\frac{d\tau}{\rho^{2}(\tau)}\Gamma(\theta(\tau; I_{1}))\right) \,.
\end{eqnarray}
In (\ref{e17}), $\theta = \theta(t; I_{1})$ is a function of $t$
as given locally by the implicit function theorem through
(\ref{e16}). The differential invariants can be used to construct
the solution for (\ref{d}). The result is
\begin{eqnarray}
A &=& \frac{\ddot\rho}{\rho}r + \frac{1}{\rho^3}\exp(\int^{t}\frac{\Gamma(\tau)}{\rho^{2}(\tau)}d\tau)\,\tilde{A}(I_{1},I_{2}) + \frac{1}{r^3}\exp(4\int^{t}\frac{\Gamma(\tau)}{\rho^{2}(\tau)}d\tau) \times \nonumber \\ \label{e18}
&\times& \int^{t}\frac{d\mu}{\rho^{2}(\mu)}\,(B^{2}\Gamma'' + BB'\Gamma')(\mu)\exp(-4\int^{\mu}\frac{\Gamma(\nu)}{\rho^{2}(\nu)}d\nu) \,,
\end{eqnarray}
where $\theta$, in the integrals, is taken as a function of $t$
through (\ref{e16}) and the implicit function theorem. $\tilde{A}$
is an arbitrary function of the differential invariants $I_1$ and
$I_2$.

To summarize, for any function $B$, the reduced Ermakov system
(\ref{red1}-\ref{red2}) has a Lie point symmetry with generator
(\ref{e13}) with $\rho \neq 0$ and $S \neq 0$ provided $A$ can be
cast in the form (\ref{e18}). While $B$ remains completely
arbitrary, $A$ belongs to a large class of functions, including
the arbitrary functions $\rho$, $\Gamma$ and $\tilde{A}$ and the
numerical constant $\kappa$. Notice that the Ermakov invariant
enters as a parameter in the symmetry generator as well as in the
function $A$. This is no surprise since the Ermakov invariant was
used to eliminate $\dot\theta$ from the equations of motion.

\subsection{The $\rho \neq 0$, $S = 0$ Case}
Now the differential invariants for $U$ are
\begin{equation}
\label{e19}
I_1 = \frac{r}{\rho} \,, \quad I_2 = \theta \,,
\end{equation}
and the corresponding solution for (\ref{d}) is
\begin{equation}
\label{e20}
A = \frac{\ddot\rho}{\rho}r + \frac{1}{r^3}\tilde{A}(\frac{r}{\rho},\theta) \,,
\end{equation}
where $\tilde{A}$ is an arbitrary function of the indicated
arguments. This class of solutions contains the arbitrary
functions $\rho$ and $\tilde{A}$, subject to $\rho \neq 0$. Also,
$B$ remains completely free and does not appear in the generator.

\subsection{The $\rho = 0$, $\Gamma \neq 0$ Case}
The differential invariants are
\begin{equation}
\label{e21}
I_1 = \left(\kappa - 2\int^{\theta}d\phi\,\frac{\Gamma(\phi)}{B(\phi)}\right)^{1/2}\,r \,, \quad I_2 = t \,,
\end{equation}
while
\begin{eqnarray}
\label{e22}
A &=& \frac{\kappa B(\theta)}{r^3}\left(\kappa - 2\int^{\theta}d\phi\,\frac{\Gamma(\phi)}{B(\phi)}\right)^{-2}\,\Gamma'(\theta) + r\,\tilde{A}(I_{1},I_{2}) \\
&-& \frac{2}{r^3}\left(\kappa - 2\int^{\theta}d\phi\,\frac{\Gamma(\phi)}{B(\phi)}\right)^{-2}\int^{\theta}d\mu\,(B\Gamma'' + B'\Gamma')(\mu)\int^{\mu}d\nu\frac{\Gamma(\nu)}{B(\nu)} \,, \nonumber
\end{eqnarray}
for $\tilde{A}$ an arbitrary function depending on the
differential invariants of the symmetry generators. Now there are
the free functions $\Gamma$, $B$ (which appears in the symmetry
generator) and $\tilde{A}$, besides the numerical constant
$\kappa$.

\subsection{The $\rho = \Gamma = 0$, $\kappa \neq 0$ Case}
This is the most simple situation. The differential invariants are $r$ and $t$ and the solution for (\ref{d}) is $A = A(r,t)$. The symmetry simply reflects the fact that $\theta$-independent equations are invariant under rotations.

\section{Connection with the $SL(2,R)$ Group}
$SL(2,R)$ is the Lie point symmetry group of non reduced Ermakov
systems \cite{leach}-\cite{goedert}, and we have to investigate
the role of this transformations group for the reduced Ermakov
systems. For simplicity, in this Section we consider Ermakov
systems containing frequencies depending only on time. In this
case (see equation (\ref{e4})) $A$ is given by
\begin{equation}
\label{e23}
A(r,\theta,t) = - \Omega^{2}(t)r + \frac{1}{r^3}(F(\theta) + B^{2}(\theta)) \,,
\end{equation}
for a time-dependent frequency $\Omega(t)$. Inserting (\ref{e23}) in the symmetry condition (\ref{d}), we get
\begin{eqnarray}
\label{e24} (\rho{\buildrel\cdots\over\rho} &+& 3\dot\rho\ddot\rho
+ 4\Omega^{2}\rho\dot\rho + 2\Omega\dot\Omega\rho^2)\,r + +
(B^{2}\Gamma'' + BB'\Gamma' + \\ &+& 4(F + B^{2})\,\Gamma - (F' +
2BB')\,S)r^{-3} = 0 \,, \nonumber
\end{eqnarray}
for $S$ defined in (\ref{e11}). Because (\ref{e24}) has to be
satisfied for arbitrary $r$, it can be split in two parts, one
corresponding to $r$, the other to $r^{-3}$,
\begin{eqnarray}
\label{e25}
\rho{\buildrel\cdots\over\rho} + 3\dot\rho\ddot\rho + 4\Omega^{2}\rho\dot\rho + 2\Omega\dot\Omega\rho^2 &=& 0 \,, \\
\label{e26}
B^{2}\Gamma'' + BB'\Gamma' +  4(F + B^{2})\,\Gamma &=& (F' + 2BB')\,S \,.
\end{eqnarray}

Equation (\ref{e26}) may be cast into a simpler form using a new independent variable $\varphi$ given by
\begin{equation}
\label{e27}
\varphi = \int^{\theta}\frac{d\mu}{B(\mu)}
\end{equation}
and a new dependent variable $W$ defined by
\begin{equation}
\label{e28} \Gamma = \frac{dW}{d\varphi}\,.
\end{equation}
Using the definition (\ref{e11}) of $S$ as well as introducing
\begin{equation}
\label{e29}
H = F + B^{2} \,,
\end{equation}
equation (\ref{e26}) is converted into
\begin{equation}
\label{e30}
\frac{d^{3}W}{d\varphi^3} + 4H\frac{dW}{d\varphi} + 2\frac{dH}{d\varphi}W = \kappa\frac{dH}{d\varphi} \,.
\end{equation}

Equations (\ref{e25}) and (\ref{e30}) can be used to search for
symmetries of specific reduced Ermakov systems. However, for {\it
arbitrary} reduced Ermakov systems, that is, for completely
arbitrary functions $H$ in (\ref{e30}), the only possibility is
\begin{equation}
\label{e31}
W = \kappa = 0 \,.
\end{equation}
This will be the choice if we are interested in symmetries valid
for {\it all} non reduced Ermakov systems, regardless this
specific form. Notice that $W = 0$ implies $\Gamma = 0$ in the
symmetry generator.

The other condition which remains is (\ref{e25}), which can be
integrated once yielding Pinney's \cite{pinney} equation,
\begin{equation}
\label{e32}
\ddot\rho + \Omega^{2}\rho = \frac{c}{\rho^3} \,,
\end{equation}
where $c$ is a constant. However, this is a nonlinear equation,
and a more fruitful approach for the study of symmetries is the
linearizing transform
\begin{equation}
\label{e33}
a = \rho^2 \,,
\end{equation}
giving
\begin{equation}
\label{e34}
{\buildrel\cdots\over{a}} + 4\Omega^{2}\dot{a} + 4\Omega\dot\Omega\,a = 0 \,.
\end{equation}
According to (\ref{e13}), the solution for this equation determines the symmetry generator
\begin{equation}
\label{e35}
U = a\frac{\partial}{\partial t} + \frac{\dot{a}r}{2}\frac{\partial}{\partial r} \,.
\end{equation}
Now, with the rescaling
\begin{equation}
\label{e36}
\alpha = a/\psi \,, \quad \tau = \int^{t}d\mu/\psi^{2}(\mu) \,,
\end{equation}
where $\psi$ is any particular solution for the time-dependent harmonic oscillator equation
\begin{equation}
\label{e37}
\ddot\psi + \Omega^{2}\psi = 0 \,,
\end{equation}
we transform (\ref{e34}) into
\begin{equation}
\label{e38}
\frac{d^{3}\alpha}{d\tau^3} = 0 \,.
\end{equation}
The general solution is (compare with (\ref{e14}))
\begin{equation}
\label{e39}
\alpha = c_0 + c_{1}\tau + c_{2}\tau^2 \,,
\end{equation}
for constants $c_0$, $c_1$ and $c_2$. Taking separately each of
these constants non zero, we obtain three symmetry generators for
arbitrary reduced Ermakov systems. In the original, non rescaled
variables and using (\ref{e35}), the corresponding symmetry
generators are
\begin{eqnarray}
\label{e40}
U_0 &=& \psi^{2}\frac{\partial}{\partial t} + \psi\dot\psi\,r\frac{\partial}{\partial r} \,,
\\
\label{e41}
U_1 &=& \psi^{2}\int^{t}\frac{d\mu}{\psi^{2}(\mu)}\,\,\,\frac{\partial}{\partial t} + (\frac{1}{2} + \psi\dot\psi\int^{t}\frac{d\mu}{\psi^{2}(\mu)})\,r\frac{\partial}{\partial r} \,, \\
\label{e42}
U_2 &=& \psi^{2}(\int^{t}\frac{d\mu}{\psi^{2}(\mu)})^{2}\,\,\,\frac{\partial}{\partial t}
+ \\
&+& (1 + \psi\dot\psi\int^{t}\frac{d\mu}{\psi^{2}(\mu)})\,
\int^{t}\frac{d\mu}{\psi^{2}(\mu)}\,r\frac{\partial}{\partial r} \,. \nonumber
\end{eqnarray}
Calculating the Lie brackets, the result is
\begin{equation}
\label{e43}
\left[U_{0},U_{1}\right] = U_0 \,, \quad \left[U_{0},U_{2}\right] = 2 U_1 \,, \quad \left[U_{1},U_{2}\right] = U_2 \,,
\end{equation}
which is the $sl(2,R)$ algebra. This shows that the symmetry group
for arbitrary reduced Ermakov systems is $SL(2,R)$. It is
interesting to note that the algebra of the vector fields $U_0$,
$U_1$ and $U_3$ is $sl(2,R)$ regardless the form of $\psi$ (does
not need to be a solution of a time-dependent harmonic
oscillator). In addition, we notice that $U_0$, $U_1$ and $U_2$ do
not depend on the Ermakov invariant, being generators of {\it
point} transformations also in the non reduced space.

\section{An Illustrative Example}
Let us consider a non reduced Ermakov system with
\begin{equation}
\label{e44}
G(\theta) = - \frac{L^{2}\sin\theta}{\cos^{3}\theta} \,, \quad L = constant \,,
\end{equation}
in equation (\ref{e2}), producing an Ermakov invariant
\begin{equation}
\label{e}
I = \frac{1}{2}(r^{2}\dot\theta)^2 + \frac{L^2}{2\cos^{2}\theta}
\end{equation}
and a reduced Ermakov system with
\begin{equation}
\label{e45}
B(\theta) = \frac{\sqrt{2I}}{\cos\theta}(1 - \frac{L^2}{2I} - \sin^{2}\theta)^{1/2} \,.
\end{equation}
We look for functions $A(r,\theta,t)$ in the reduced Ermakov
system that lead to Lie point symmetries as described by the class
of solutions in subsection 3.1. Other classes of solutions may
also be studied, but we restrict ourselves to this case in view of
its generality. In fact, the solutions described in 3.1 have too
many arbitrary functions, and we restrict ourselves to the choices
\begin{equation}
\label{e46}
\kappa = 0 \,, \quad \Gamma(\theta) = \Gamma_{0}\sin\theta \,, \quad \rho(t) = \cos\omega_{0}t \,,
\end{equation}
for constants $\Gamma_0$ and $\omega_0$.

In order to explicitly write the solution $A$, we have first to
obtain the differential invariants (\ref{e16}-\ref{e17}) of the
Lie symmetry. According to (\ref{e11}) and (\ref{e45}-\ref{e46}),
we have
\begin{equation}
\label{e47}
S(\theta) = \frac{2\Gamma_0}{\cos\theta}(1 - \frac{L^2}{2I} - \sin^{2}\theta) \,,
\end{equation}
which, substituted in (\ref{e16}), gives the differential invariant
\begin{equation}
\label{e48}
I_1 = \frac{1}{2\Gamma_0}\frac{1}{\sqrt{1-L^{2}/2I}}\,{\rm arctanh}\left(\frac{\sin\theta}{\sqrt{1-L^{2}/2I}}\right) - \frac{\tan\omega_{0}t}{\omega_0} \,.
\end{equation}
To obtain the second differential invariant from (\ref{e17}), we
have to solve (\ref{e48}) for $\theta$, which in this case yields
\begin{equation}
\label{e49}
\theta = \arcsin\left(\sqrt{1-\frac{L^2}{2I}}\tanh\left(2\Gamma_{0}\sqrt{1-\frac{L^2}{2I}}(\frac{\tan\omega_{0}t}{\omega_0} + I_{1})\right)\right) \,.
\end{equation}
With this result and performing the necessary quadratures, we find
from (\ref{e17}) the differential invariant
\begin{equation}
\label{e50}
I_2 = (1 - \frac{\sin^{2}\theta}{1-L^{2}/2I})^{1/4}\,\frac{r}{\cos\omega_{0}t} \,.
\end{equation}
Again using (\ref{e49}) and performing the necessary integrations
in (\ref{e18}), we find  the set of admissible $A$ functions given
by
\begin{equation}
\label{e51}
A = - \omega_{0}^{2}r - \frac{(1 - L^{2}/2I)\sin^{2}\theta}{(1 - L^{2}/2I - \sin^{2}\theta)}\,\frac{I}{r^3} + \frac{r}{\cos^{4}\omega_{0}t}\,\tilde{A}(I_{1},I_{2}) \,,
\end{equation}
for arbitrary $\tilde{A}$ depending on the differential invariants
at (\ref{e48}) and (\ref{e50}). We have redefined the arbitrary
function $\tilde{A}$ in (\ref{e18}) by making $\tilde{A}
\rightarrow \tilde{A}/I_{2}$ for the sake of a better notation.

To summarize, reduced Ermakov systems (\ref{red1}-\ref{red2}) with
$A$ given by (\ref{e51}) and $B$ given by (\ref{e45}) do possess
Lie point symmetries with generator
\begin{eqnarray}
\label{e52}
U &=& (-\omega_{0}\sin\omega_{0}t\cos\omega_{0}t + \Gamma_{0}\sin\theta)\,r\frac{\partial}{\partial r} + \\ &+& \frac{2\Gamma_0}{\cos\theta}(1 - \frac{L^2}{2I} - \sin^{2}\theta)\,\frac{\partial}{\partial\theta}   + \cos^{2}\omega_{0}t\,\frac{\partial}{\partial t} \,, \nonumber
\end{eqnarray}
as follows from (\ref{e13}) and (\ref{e46}-\ref{e47}). A remark
applicable here and in most of other cases is that the Ermakov
invariant does appear in the generator of symmetries. This is no
surprise since the reduced Ermakov system was written after
restricting the trajectories to the level surfaces of the Ermakov
invariant. Therefore, the reduced Ermakov systems contains $I$ as
a parameter. Hence, (\ref{e52}) is the generator of a point
transformation on the reduced space and of a dynamical
transformation on the non reduced  space, where the Ermakov
invariant is written as a function of $r$, $\theta$ and
$\dot\theta$ as in (\ref{e}).

\section{Application to the Generalized Paul Trap}
In practical applications, more often one does not assume the form
of the generator and try to obtain the equations of motion for
which it is a symmetry, like in Section 5. Instead, one has some
specific equations of motion and then look for symmetries. This is
the approach we follow in this Section, searching for Lie point
symmetries for the following class of Ermakov systems, written in
cartesian coordinates:
\begin{eqnarray}
\label{e53}
\ddot{x} + (\Omega^{2}(t) - \frac{\varsigma(x\dot{y}-y\dot{x},y/x)}{(x^2 + y^{2})^{3/2}})\,x &=& \frac{L^2}{x^3} \,,\\
\label{e54}
\ddot{y} + (\Omega^{2}(t) - \frac{\varsigma(x\dot{y}-y\dot{x},y/x)}{(x^2 + y^{2})^{3/2}})\,y &=& 0 \,,
\end{eqnarray}
where $\Omega$ and $\varsigma$ are initially arbitrary functions
of the indicated arguments, and $L$ is a constant. For constant
$\Omega$ and $\varsigma$, these are the equations of motion for an
ion in the presence of a Paul trap \cite{paul} with equal secular
frequencies. In this context, we call (\ref{e53}-\ref{e54}) the
equations of motion for a generalized Paul trap. Paul traps are a
standard configuration used in ion trapping experiments
\cite{baumann}. Also notice that, for $\varsigma$ depending only
on $y/x$, equations (\ref{e53}-\ref{e54}) is a particular case of
the Kepler-Ermakov systems, which are linearizable through point
transformations \cite{haasg, athornel}. Here we ask for the
classes of functions $\Omega$ and $\varsigma$ for which the
corresponding reduced Ermakov systems do admit Lie point
symmetries.

In polar coordinates, the generalized Paul trap equations can be cast in the Ermakov form (\ref{e1}-\ref{e2}) with
\begin{equation}
\label{e55}
F(\theta) = \frac{L^2}{\cos^{2}\theta} \,, \quad G(\theta) = - \frac{L^{2}\sin\theta}{\cos^{3}\theta} \,, \quad \omega^2 = \Omega^{2}(t) - \frac{\varsigma(r^{2}\dot\theta,\tan\theta)}{r^3} \,.
\end{equation}
Notice the generalized character of $\omega$, which is not necessarily a function of time only.

The associated Ermakov invariant is
\begin{equation}
\label{e56}
I = \frac{1}{2}(r^{2}\dot\theta)^2 + \frac{L^2}{2\cos^{2}\theta}
\end{equation}
and the reduced Ermakov system is constructed with the functions
\begin{eqnarray}
\label{e57}
A(r,\theta,t) &=& - \Omega^{2}(t)r + \frac{\sigma(\theta)}{r^2} + \frac{2I}{r^3} \,,\\
\label{e58}
B(\theta) &=& \frac{\sqrt{2I}}{\cos\theta}(1 - \frac{L^2}{2I} - \sin^{2}\theta)^{1/2} \,,
\end{eqnarray}
where we have defined
\begin{equation}
\label{e59}
\sigma(\theta) = \varsigma(B(\theta),\tan\theta) \,.
\end{equation}

Let us search for symmetries of the type shown in Section 3. For
this, a convenient approach is to substitute $A$ and $B$ in the
symmetry condition (\ref{d}) looking for some symmetry generator.
The symmetry condition (\ref{d}) then gives
\begin{eqnarray}
\label{e60}
(\rho{\buildrel\cdots\over\rho} &+& 3\dot\rho\ddot\rho + 4\Omega^{2}\rho\dot\rho + 2\Omega\dot\Omega\rho^2)\,r + (-S(\theta)\sigma' + (3\Gamma(\theta)-\rho\dot\rho)\sigma)r^{-2}  \\
&+& (\frac{2I(1-L^{2}/2I-\sin^{2}\theta)}{\cos^{2}\theta}\,\Gamma'' - \frac{L^{2}\sin\theta}{\cos^{3}\theta}\,\Gamma' + 8I\,\Gamma)\,r^{-3} = 0 \,, \nonumber
\end{eqnarray}
where $S(\theta)$ depends on $\Gamma(\theta)$ according to
(\ref{e11}). Compare (\ref{e60}) with (\ref{e24}). There are
similarities since $B(\theta)$ here is the same as in Section 5,
but now there is a contribution proportional to $r^{-2}$, peculiar
to Kepler-Ermakov systems.

As in Section 5, we split the symmetry condition (\ref{e60}) into three equations, corresponding to different powers of $r$,
\begin{eqnarray}
\label{e61}
\rho{\buildrel\cdots\over\rho} + 3\dot\rho\ddot\rho + 4\Omega^{2}\rho\dot\rho + 2\Omega\dot\Omega\rho^2 &=& 0 \,, \\
\label{e62}
S(\theta)\sigma' + (\rho\dot\rho -3\Gamma(\theta))\sigma &=& 0 \,, \\
\label{e63}
\frac{2I(1-L^{2}/2I-\sin^{2}\theta)}{\cos^{2}\theta}\,\Gamma'' &-& \frac{L^{2}\sin\theta}{\cos^{3}\theta}\,\Gamma' + 8I\,\Gamma = 0 \,.
\end{eqnarray}
For consistency with equation (\ref{e59}), in equation
(\ref{e62}) we must have
\begin{equation}
\label{e64}
\rho\dot\rho = \frac{\Omega_{0}}{2} \,,
\end{equation}
for a constant $\Omega_0$. Integrating, we get
\begin{equation}
\label{e65}
\rho = (\rho_{0}^2 + 2\Omega_{0}t)^{1/2} \,,
\end{equation}
where $\rho_0$ is a constant. Inserting this into (\ref{e61}), we
obtain the following class of frequencies $\Omega$,
\begin{equation}
\label{e66}
\Omega = \frac{\Omega_0}{\rho_{0}^2 + 2\Omega_{0}t} \,.
\end{equation}

Equation (\ref{e63}) can be best handled using the new independent
variable
\begin{equation}
\label{e67}
\varphi = \int^{\theta}\frac{d\mu}{B(\mu)} = \frac{1}{\sqrt{2I}}\arcsin(\frac{\sin\theta}{\sqrt{1-L^{2}/2I}}) \,,
\end{equation}
yielding
\begin{equation}
\label{e68}
\frac{d^{2}\Gamma}{d\varphi^2} + 8I\Gamma = 0 \,,
\end{equation}
with solution
\begin{equation}
\label{e69}
\Gamma = \Gamma_{1}\cos(2\sqrt{2I}\varphi) + \Gamma_{2}\sin(2\sqrt{2I}\varphi) \,,
\end{equation}
where $\Gamma_1$ and $\Gamma_2$ are constants.

There remains the equation (\ref{e62}). Calculating $S(\theta)$ via (\ref{e11}) and using the new independent variable $\varphi$, this equation reads
\begin{equation}
\label{e70}
(\kappa + \frac{\Gamma'}{4I})\,\frac{d\sigma}{d\varphi} + (\Omega_{0} - 3\Gamma)\,\sigma = 0 \,,
\end{equation}
with $\Gamma$ as in (\ref{e69}).

There is no difficulty at solving (\ref{e70}) in the general case,
but the solution is somewhat complicated. It seems more
illustrative to obtain the general solution in some specific
cases, listed bellow.

\subsection{The $\kappa \neq 0$, $\Gamma_1 = \Gamma_2 = 0$ Case}
The general solution for (\ref{e70}) is
\begin{equation}
\label{e71}
\sigma = \sigma_{0}e^{-\frac{\Omega_{0}\varphi}{\kappa}} = \sigma_{0}\exp(-\frac{\Omega_{0}}{\kappa\sqrt{2I}}\arcsin(\frac{\sin\theta}{\sqrt{1-L^{2}/2I}})) \,,
\end{equation}
for any constant $\sigma_0$.

\subsection{The $\Gamma_{1}^2 + \Gamma_{2}^2 \neq 0$, $\kappa = \Omega_0 = 0$ Case}
In this situation the general solution is
\begin{equation}
\label{e72}
\sigma = \sigma_0\,\left(\frac{d\Gamma}{d\varphi}\right)^{-3/2} \,,
\end{equation}
where $\sigma_0$ is a constant and $\Gamma$ is given by (\ref{e69}), with $\varphi$ defined in (\ref{e67}).

In both subcases (\ref{e71}) and (\ref{e72}), the solution
contains the Ermakov invariant (\ref{e56}), which is dependent on
$r^{2}\dot\theta$. Hence, at least in the generalized Paul trap
case, we have not found solutions for which $\sigma$ is a function
of $\theta$ only, characterizing a Kepler-Ermakov system.

To summarize, we found the functions $\sigma$ (alternatively,
$\varsigma$) and $\Omega$ so that the reduced Ermakov system for
the generalized Paul trap equations do possess Lie point
symmetries. The functions $\sigma$ do satisfy (\ref{e70}), as in
subsections 6.1 and 6.2, while the frequency $\Omega$ is of the
form (\ref{e66}). The corresponding generator of symmetry is build
with $\rho$ given by (\ref{e65}) and $\Gamma$ given by
(\ref{e69}).

\section{Conclusion}
We presented the general treatment for Lie point symmetries of
reduced Ermakov systems. We found four classes of reduced Ermakov
systems possessing Lie point symmetries, all of them involving
arbitrary functions. We have applied the results to two different
systems, including a generalized Paul trap. From the theoretical
viewpoint, the most important result we have found is the fact
that the $SL(2,R)$ group is more exactly a property of the {\it
reduced} Ermakov system. For Lagrangian Ermakov systems, the
existence of the Ermakov invariant follows from a dynamical
symmetry. Then, for reduced Lagrangian Ermakov systems, the
symmetry structure can be separated in two distinct parts: a
dynamical symmetry leading to the Ermakov invariant, and the
$SL(2,R)$ group for the reduced Ermakov system, the later a
consequence of the first symmetry.


\begin{thebibliography}{99}
\bibitem{ermakov} Ermakov V. P. 1880 {\it Univ. Izv. Kiev} {\bf 20} 1.
\bibitem{ray} Ray J. R. and Reid J. L. 1979 {\it Phys. Lett. A} {\bf 71} 317.
\bibitem{lewis} Lewis H. R. 1967 {\it Phys. Rev. Lett.} {\bf 18} 510.
\bibitem{haasPRA} Haas F. 2002 {\it Phys. Rev. A} {\bf 65} 033603.
\bibitem{lidsey} Lidsey J. E. 2004 {\it Class. and Quantum Gravity} {\bf 21} 777.
\bibitem{hawkins} Hawkins R. M. and Lidsey J. E. 2002 {\it Phys. Rev. D} {\bf 66} 023523.
\bibitem{ioffe} Ioffe M. V. and Korsch H. J. 2003 {\it Phys. Lett. A} {\bf 311} 200.
\bibitem{rosu} Rosu H. C. 2002 {\it Physica Scripta} {\bf 65} 296.
\bibitem{akulov} Akulov V. P, Catto S., Cebio$\rm{\check{g}}$lu O. and Pashnev A. 2003 {\it Phys. Lett. B} {\bf 575} 137.
\bibitem{haasJPA} Haas F. 2002 {\it J. Phys. A: Math. Gen.} {\bf 35} 2925.
\bibitem{reid} Reid J. L. and Ray J. R. 1980 {\it J. Math. Phys.} {\bf 21} 1583.
\bibitem{athornelinear} Athorne C., Rogers C., Ramgulam U. and Osbaldestin A. 1990 {\it Phys. Lett. A} {\bf 143} 207.
\bibitem{haasg} Haas F. and Goedert J. 1999 {\it J. Phys. A: Math. Gen.} {\bf 32} 2835.
\bibitem{olver} Olver P. J. 1986 {\it Applications of Lie Groups to Differential Equations} (Springer-Verlag: Berlin).
\bibitem{leach} Leach P. G. L. 1991 {\it Phys. Lett. A} {\bf 158} 102.
\bibitem{gov1} Govinder K. S., Athorne C. and Leach P. G. L. 1993 {\it J. Phys. A: Math. Gen.} {\bf 26} 4035.
\bibitem{gov2} Govinder K. S. and Leach P. G. L. 1994 {\it J. Phys. A: Math. Gen.} {\bf 27}  4153.
\bibitem{gov3} Govinder K. S. and Leach P. G. L. 1994 {\it Phys. Lett. A} {\bf 186} 391.
\bibitem{goedert} Goedert J. and Haas F. 1998 {\it Phys. Lett. A} {\bf 239} 348.
\bibitem{haasPLA} Haas F. and Goedert J. 2001 {\it Phys. Lett. A} {\bf 279} 181.
\bibitem{karasu} Karasu A. and Yildirim H. 2002 {\it J. Nonlinear Math. Phys.} {\bf 9} 475.
\bibitem{athornel} Athorne C. 1991 {\it J. Phys. A: Math. Gen.} {\bf 24} L1385.
\bibitem{pinney} Pinney E. 1950 {\it Proc. Am. Math. Soc.} {\bf 1} 681.
\bibitem{paul} Paul W. and Steinwedel H. 1953 {\it Z. Naturforschg.} {\bf 8a} 448.
\bibitem{baumann} Baumman G. and Noemacher T. F. 1992 {\it Phys. Rev. A} {\bf 46} 2682.
\end{thebibliography}
\end{document}